\documentclass[twocolumn,twocolappendix,tighten,times,astrosymb]{aastex631}

\usepackage{amsmath}
\usepackage{soul}
\usepackage{microtype}
\usepackage[english]{babel}
\usepackage{cancel}
\usepackage{soul}
\hypersetup{linkcolor=blue,citecolor=blue,filecolor=blue,urlcolor=blue}

\newcommand{\be}{\begin{equation}}
\newcommand{\ee}{\end{equation}}

\long\def\exclude#1{}

\shorttitle{Turbulent corona as the source of neutrinos from NGC 1068}
\shortauthors{}
\begin{document}

\title{\large A magnetized strongly turbulent corona as the source of neutrinos from NGC 1068
}

\correspondingauthor{damiano.fiorillo@nbi.ku.dk }
\author[0000-0003-4927-9850]{Damiano F. G. Fiorillo}
\affiliation{Niels Bohr International Academy, Niels Bohr Institute, University of Copenhagen, 2100 Copenhagen, Denmark}
\author[0000-0001-8822-8031]{Luca Comisso}
\affiliation{Department of Astronomy and Columbia Astrophysics 
Laboratory, Columbia University, New York, NY 10027, USA}
\author[0000-0003-0543-0467]{Enrico Peretti}
\affiliation{Université Paris Cité, CNRS, Astroparticule et Cosmologie, 10 Rue Alice Domon et Léonie Duquet, F-75013 Paris, France}
\author[0000-0001-6640-0179]{Maria Petropoulou}
\affiliation{Department of Physics, National and Kapodistrian University of Athens, University Campus Zografos, GR 15784, Athens, Greece }
\affiliation{Institute of Accelerating Systems \& Applications, University Campus Zografos, Athens, Greece}
\author[0000-0002-1227-2754]{Lorenzo Sironi}
\affiliation{Department of Astronomy and Columbia Astrophysics 
Laboratory, Columbia University, New York, NY 10027, USA}
\affiliation{Center for Computational Astrophysics, Flatiron Institute, 162 5th Avenue, New York, NY 10010, USA}

\begin{abstract}
The cores of active galactic nuclei (AGN) are potential accelerators of 10--100~TeV cosmic rays, in turn producing high-energy neutrinos. This picture was confirmed by the compelling evidence of a TeV~neutrino signal from the nearby active galaxy NGC~1068, leaving open the question of which is the site and mechanism of cosmic ray acceleration. One candidate is the magnetized turbulence surrounding the central supermassive black hole. Recent particle-in-cell simulations of magnetized turbulence indicate that stochastic cosmic ray acceleration is non-resonant, in contrast to the assumptions of previous studies. We show that this has important consequences on a self-consistent theory of neutrino production in the corona, 
leading to a more rapid cosmic ray acceleration than previously considered. 
The turbulent magnetic field fluctuations needed to explain the neutrino signal are consistent with a magnetically powered corona. We find that strong turbulence, with turbulent magnetic energy density higher than $1\%$ of the rest mass energy density, naturally explains the normalization of the IceCube neutrino flux, in addition to the neutrino spectral shape. 
Only a fraction of the protons in the corona, which can be directly inferred from the neutrino signal, are accelerated to high energies. Thus, in this framework, the neutrino signal from NGC~1068 provides a testbed for particle acceleration in  magnetized turbulence.
\end{abstract}


\section{Introduction}

Active galactic nuclei (AGN), namely the compact regions at the center of active galaxies, are powerful emitters of electromagnetic radiation across a broad range of wavelengths. The source of their power is associated with the accretion of matter onto a central supermassive black hole (SMBH). AGN are observed throughout the whole electromagnetic spectrum, from radio up to hard X-rays and often also in gamma rays. The spectral components common to all non-jetted AGN (for jetted AGN, the jet emission might hide these components) are (see, e.g.,~\cite{2017A&ARv..25....2P}): (1) an optical thermal peak emitted by the accretion disk, (2) a second thermal peak in the infrared associated to a parsec-scale dusty torus, and (3) an X-ray power-law spectrum with an photon index close to -2 associated with a region somehow close to the SMBH, often referred to as the AGN corona. The size and precise location of the corona relative to the SMBH and the accretion disk are matter of debate in the scientific community~\citep[see e.g.][and references therein]{Cackett2021}. Nevertheless, it is generally believed that the corona could extend from a few up to hundreds of gravitational radii.

The large amount of power emitted by AGN in the form of electromagnetic radiation has led to their widespread consideration as candidates for the acceleration of high-energy cosmic rays \citep{Berezinsky_1977,Berezinsky_1981,Silberberg_1979,Eichler_1979}. This possibility has recently come to the forefront after the IceCube collaboration reported the measurement, at a $4.2\;\sigma$ level of significance, of a high-energy neutrino flux in the 1.5--15~TeV energy range from the active galaxy NGC~1068~\citep[][]{Aartsen_Icecube_2020,IceCube-NGC1068}. The detection of high-energy neutrinos is usually interpreted as the smoking-gun signature of the interaction of hadronic cosmic rays with matter or radiation; indeed, recent work by \cite{Das:2024vug} has ruled out, in a nearly model-independent way, the possibility that the neutrinos could have originated from leptonic or beta decay processes. Therefore, the IceCube measurement requires hadronic cosmic rays accelerated up to tens of TeV. On the other hand, the absence of a comparable TeV gamma-ray flux~\citep{MAGIC-UL-NGC1068} implies that these neutrinos must be produced in a region optically thick to $\gamma\gamma$ pair production. Such a region could be located close to the central black hole, where the optical and X-ray fields would absorb the gamma-rays from neutral pion decays (associated with neutrinos) and reprocess them to lower energies~\citep{Berezinsky_1981}. Thus, the GeV gamma-rays observed by Fermi-LAT~\citep{Abdo2010} would be reasonably produced in a different, outer region, e.g., a weak jet~\citep{Lenain2010}, the circumnuclear starburst region~\citep{Yoast-Hull2013,Ambrosone:2021aaw,Eichmann_2022}, a large scale AGN-driven outflow~\citep{Lamastra_2016}, failed line-driven winds~\citep{Inoue_S_2022}, or an ultra-fast outflow~\citep{Peretti_2023}. On the other hand, neutrino emission is strongly inferred to be associated with the corona \citep[see][for a comprehensive review]{Padovani2024}.

The main question that remains open is what mechanism would accelerate protons in the corona up to the required energies of tens of TeV. Various possibilities have been explored in the literature. In the case of diffusive shock acceleration \citep{Inoue_2020}, or gyroresonant stochastic acceleration \citep{Murase:2019vdl}, proton acceleration to high energies is limited by photohadronic cooling, which sets the energy scale for the observed IceCube signal. \cite{Mbarek:2023yeq} considered proton re-acceleration in turbulence after a pre-acceleration phase in intermittent reconnection layers in the vicinity of the black hole as a possible explanation for the neutrino signal. Magnetic reconnection was considered in~\cite{Kheirandish_2021}, modeled as a power-law injection up to the maximum energies allowed by the acceleration, leading to the conclusion that it was disfavored as an explanation for the neutrinos because of the large number of events expected above tens of TeV. On the other hand, in our previous work~\citep{Fiorillo:2023dts}, we showed that a reconnection layer might be entirely responsible for the proton acceleration. This scenario is markedly distinct from the previous ones, in that individual protons could be accelerated to energies much higher than hundreds of TeV without efficient cooling, but the maximum global energy density that the proton population can reach is limited by the available magnetic energy density, which determines the energy scale of the peak in the IceCube neutrino spectrum.

In this work, we revisit the scenario of stationary turbulent acceleration proposed in~\cite{Murase:2019vdl} incorporating recent advancements from particle-in-cell (PIC) simulations in describing the particle acceleration process. Specifically, particle acceleration is governed by non-resonant interactions with turbulent fluctuations~\citep{CS19,2020ApJ...893L...7W,Zhdankin20,2022PhRvD.106b3028B}, resulting in an energy-independent timescale for the particle acceleration process. Furthermore, we explicitly account for the fact that the volumetric rate of proton energization is bounded by the rate of turbulent magnetic energy dissipation. Adopting these constraints, we find that previously considered scenarios are difficult to reconcile with the high neutrino flux measured by IceCube. 
In this work, on the other hand, we propose a physically motivated model involving non-resonant proton acceleration in strong turbulence, showing that it is consistent with both the normalization and spectral shape of the IceCube signal.

\section{Turbulent model: Coronal properties}\label{sec:coronal_properties}

Before providing a detailed description of the proton and neutrino spectral features inside the corona, in this section we outline the physical parameters of the turbulent coronal model using order-of-magnitude arguments.

In this model, the corona is a spherical region with radius $R$, with a magnetic field of strength $B$ and stationary turbulence, surrounding a black hole with mass $M=M_7\; 10^7\;M_\odot$, consistent with the order of magnitude in~\cite{Padovani:2024ibi}. The gravitational radius of the black hole is $r_g=G M/c^2\simeq1.5\times 10^{12}\;M_7$~cm, where $G$ is the gravitational constant and $c$ is the speed of light.
The corona contains a population of leptons, with number density $n_e$, and protons, with number density $n_p$. The number density of leptons can be directly inferred by the Compton opacity of the corona, whose characteristic values are in the ballpark of $\tau_T\simeq 0.5$~\citep[e.g.][]{2018MNRAS.480.1819R}. This implies a lepton number density
\begin{equation}\label{eq:thomson}
    n_e \simeq \frac{\tau_T}{\sigma_T R}\simeq  2.5\times 10^{10}\;\frac{20 r_g}{M_7 R}\; \mathrm{cm}^{-3},
\end{equation}
where $\sigma_T$ is the Thomson cross section.

In our previously proposed reconnection model \citep{Fiorillo:2023dts} we assumed that the coronal plasma was predominantly composed of leptons, with $n_e\gg n_p$. The rationale is that, if reconnecting current sheets form in the magnetospheric region, the bulk of the protons are expected to be accelerated at a few $r_g$ from the event horizon. In order to explain the neutrino signal, a proton density much smaller than the lepton density estimated from Eq.~(\ref{eq:thomson}) is required, thus leading to the conclusion that the corona must be pair-dominated within a reconnection-driven model. On the other hand, in a turbulence-driven model, which is not generally associated with the black hole magnetosphere but rather with the accretion flow, one does not necessarily expect the bulk of the protons to be accelerated; in other words, the proton number density estimated from the neutrino signal refers to the number density of non-thermal protons, $n_p^\mathrm{nt}$. The latter might be a small fraction of the total amount of protons in the corona, $n_p^\mathrm{nt}\ll n_p$, but there is no fixed prediction as to what fraction. Thus, the neutrino signal in this case does not allow us to uniquely infer the proton and pair composition of the corona, and one could still have a system with $n_p \simeq n_e$. We maintain this assumption throughout the subsequent discussion.

The proton number density impacts the properties of the corona, since it affects the magnetization of the plasma, defined as  
\begin{equation} \label{def_sigma}
    \sigma=\frac{B^2}{4\pi(n_e m_e+n_p m_p)c^2} \, ,
\end{equation}
where $B$ is the total magnetic field strength. With our assumption that $n_p \simeq n_e$, protons dominate the mass density and directly determine the plasma magnetization $\sigma$. 
The strength of the turbulent magnetic field fluctuations, characterized by the root-mean-squared value of the turbulent magnetic field, $\delta B$, is generally smaller than the total magnetic field strength. Therefore, we denote 
\begin{equation}
    \sigma_{\rm tur}=\sigma\eta_B\, , \quad \eta_B= \left( \frac{\delta B}{B} \right)^2 \, , 
\end{equation}
as the magnetization associated with the turbulent field alone. With this definition, we have $(v_A/c)^2=\sigma_{\rm tur}/(1+\sigma_{\rm tur})\simeq \mathrm{min}(\sigma_{\rm tur},1)$, where $v_A$ indicates the Alfv{\'e}n velocity associated with the turbulent component of the magnetic field. We can also relate the total plasma magnetization $\sigma$ to the commonly used proton plasma beta, $\beta=8\pi n_p k_B T_p/B^2$. As in~\cite{Murase:2019vdl}, assuming the virial temperature $k_B T_p=G M m_p/3 R$, allows us to express the proton plasma beta as 
\begin{equation}
    \beta = \frac{2k_B T_p}{\sigma m_p c^2} = \frac{2r_g}{3 R \sigma } \, . 
\end{equation}

In our model, a fraction of the coronal protons are accelerated by scattering off turbulent fluctuations. In the historical resonant picture, particles with a gyroradius $\rho_g$ are resonantly accelerated by turbulent fluctuations of a comparable size $\lambda\sim \rho_g$. The rate of energy gain in this process is proportional to the amount of turbulent power at wavenumbers $k\sim \rho_g^{-1}$, resulting in an energy-dependent acceleration rate. However, even within the framework of quasi-linear theory, the interaction between charged particles and turbulent fluctuations, interpreted as waves, is not strongly resonant due to significant broadening by wave damping~\citep{Chandran:2000hp,Demidem:2019jzn}. 
In recent years, it has emerged that non-resonant scattering on turbulent fluctuations with size $\lambda \gtrsim \rho_g$ generally constitutes the dominant driver of particle acceleration. This argument is supported by particle-in-cell (PIC) simulations \citep{CS19,2020ApJ...893L...7W,Zhdankin20,2022PhRvD.106b3028B}, which showed that the characteristic timescale associated with stochastic particle acceleration, $t_\mathrm{acc}$, is essentially independent of particle energy and dominated by the largest scale structures that carry most of the magnetic turbulent energy.

As a typical size for the largest scale structures, we should take the turbulence coherence length (i.e., the turbulence-driving scale), $\ell$, which is taken to be a fraction of the coronal size, $\ell=\eta R$, with $\eta \leq 1$. According to PIC simulations, the diffusion coefficient in momentum space resulting from stochastic particle acceleration is \citep{CS19} $D_{p} \simeq 0.1 \sigma_{\rm tur} p^2 c/\ell$. \footnote{This result was obtained for strong turbulence (for weak wave-like turbulence, see discussion in \citet{Demidem:2019jzn}).} Therefore, as an order of magnitude for the acceleration timescale, we take
\begin{equation}\label{eq:acceleration_timescale}
    t_\mathrm{acc} \equiv \frac{p^2}{D_p} \simeq \frac{10}{\sigma_{\rm tur}} \frac{\ell}{c} \, .
\end{equation} 
 The factor $\sigma_{\rm tur}$ in the denominator accounts for the typical energization in every scattering event.

Turbulent magnetic field fluctuations play an important role also in proton confinement. For protons with gyro-radii $\rho_g \ll \ell$, scattering on intermittent small-scale field reversals becomes important in strong turbulence, which constitutes the main focus of our study. On general grounds, the proton mean free path depends on the gyro-radius as $\lambda_\mathrm{mfp} \simeq \ell \left( \rho_g/\ell \right)^{\delta}$, with $\delta > 0$, namely longer proton residence times for lower-energy protons. The exact value of $\delta$, affected by turbulence intermittency, is a subject of active investigation, though theoretical arguments and simulations suggest $0.3 \leq \delta \leq 0.5$ \citep{Lem2023JPlPh,Kempski23}. 
Here, we use $\delta = 1/3$, but varying $\delta$ within the range written above does not alter our conclusions. In an effective random-walk fashion, the residence time within the corona is $t_{\rm esc}\simeq R^2/\lambda_{\rm mfp} c$.
Considering that the minimum escape timescale is $R/c$, when angular diffusion is so slow that the particles freely stream out of the system, we write the escape timescale as
\begin{equation}
    t_\mathrm{esc}\simeq \frac{R}{c}\mathrm{max}\left[1,\frac{R}{\ell }\left(\frac{eB \ell }{E_p}\right)^{1/3}\right], 
\end{equation} 
where $E_p$ indicates the proton energy.

The dominant factor limiting the acceleration of protons up to the Hillas limit of the turbulent cascade, $E_{p,\mathrm{max}} = eB \ell$, is their cooling. The bulk of the energy injected in non-thermal protons must peak at a characteristic energy around $E_{p,\mathrm{peak}}\simeq 20$~TeV in order to explain a neutrino peak energy $E_{\nu,\mathrm{peak}}\simeq 1$~TeV. Therefore, one must assume that protons above $E_{p,\mathrm{peak}}$ are cooled faster than they are accelerated by the magnetized turbulence. 

At $E_p = 20$~TeV, there are three competing processes that cool protons, namely $pp$ scattering on the dense coronal matter, photopion scattering on the X-ray coronal field, and Bethe-Heitler (BH) scattering on the optical-ultraviolet (OUV) field produced in the accretion disk. The $pp$ energy loss timescale can be estimated as
\begin{equation}
    t_{pp}^{-1}(E_p)\simeq n_p \sigma_{pp}(E_p) \kappa_p c.
\end{equation}
Here $\kappa_p\simeq 0.5$ is the typical inelasticity in $pp$ collisions, and $\sigma_{pp}(E_p)$ is the total $pp$ cross section; for our numerical calculations, we use the expression valid above 10~GeV from~\cite{1996A&A...309..917A}.

\begin{figure}
    \centering
    \includegraphics[width=0.5\textwidth]{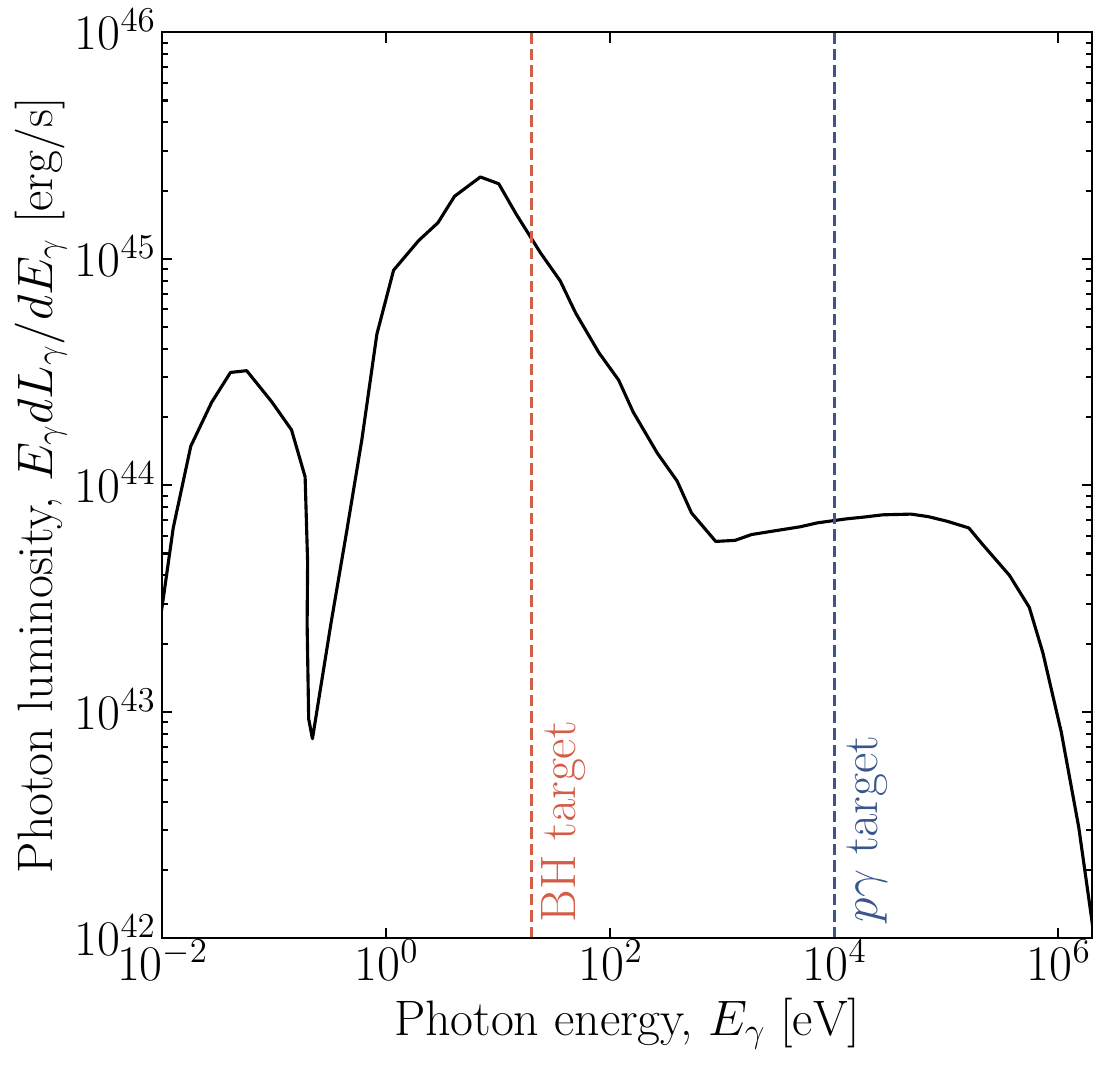}
    \caption{Electromagnetic luminosity emitted from the AGN as a function of the photon energy $E_\gamma$. We identify by dashed lines the energies corresponding to the dominant target for Bethe-Heitler (BH) and $p\gamma$ interactions for a proton with energy $E_p=20$~TeV, as identified in the text.}
    \label{fig:photon_spectrum}
\end{figure}

The cooling rates due to photohadronic interactions depend sensitively on the electromagnetic spectrum inside the corona. For this work, we use the electromagnetic spectrum shown in Fig.~\ref{fig:photon_spectrum} in the form of the photon luminosity $dL_\gamma/dE_\gamma$, defined as the amount of energy emitted by the AGN per unit time, differential in the photon energy $E_\gamma$. 
The multiwavelength AGN spectrum from optical to X-ray is inferred following \cite{Marconi_2004_SED} under the assumption of an X-ray luminosity $L_X = 10^{44} \, \rm erg \, s^{-1}$, where the X-ray luminosity is integrated from 2 keV to 10 keV. The far-infrared torus component is computed according to the prescription of \cite{Mullaney2011}.
From the photon luminosity, we extract the photon number density per unit energy
\begin{equation}
    \frac{dN_\gamma}{dV dE_\gamma}=n_\gamma(E_\gamma)=\frac{1}{4\pi R^2 c E_\gamma}\frac{dL_\gamma}{dE_\gamma}.
\end{equation}

From the photon spectrum, we can directly extract the timescales for photohadronic interactions. The first process we consider is Bethe-Heitler (BH) or pair photoproduction $p\gamma\to p e^+ e^-$. The timescale for BH energy loss can be obtained by integrating the photon spectrum over the cross section and the inelasticity for the scattering. For this calculation, we use the fits provided in~\cite{1992ApJ...400..181C}. The dominant scatterers for a proton with a typical energy $E_p=20$~TeV are photons with an approximate energy $E_\gamma\simeq m_e m_p c^4/E_p \simeq 20$~eV, which therefore lie in the high-energy tail of the big blue bump often referred to as the soft-excess.

The second process we consider is inelastic photopion production, often called simply $p\gamma$ interaction, which usually leads to the production of pions in the final state. At relatively low center-of-mass energies, the interaction is dominated by intermediate resonances, the most prominent of which is the $\Delta^+$ resonance, while at higher center-of-mass energies it is dominated by deep inelastic scattering leading to multi-pion production. We model the energy loss rate for $p\gamma$ scattering following~\cite{Atoyan:2002gu} and~\cite{Dermer:2003zv}. In this case, the center-of-mass energy for the scattering is typically higher than in BH interactions, due to the higher threshold for pion production, so for a proton with energy $E_p=20$~TeV the dominant target photons have an energy $E_\gamma\simeq \overline{\epsilon}m_e m_p c^4/E_p^2\simeq 10$~keV, corresponding to the X-ray spectrum originating in the AGN corona, with $\overline{\epsilon}=390$. 

For completeness, we also account for synchrotron energy losses  \footnote{We consider proton acceleration to $E_p \gg \sigma_{\rm tur} m_p c^2$, so the effect of pitch angle anisotropy is negligible \citep{CS19,ComissoApJL2020}.}
\begin{equation}
t_\mathrm{synch}^{-1}=
    \frac{2\sigma_T m_e^2 E_p \sigma n_e}{3m_p^3 c},
\end{equation}
although these are strongly subdominant compared to all the other cooling channels.

We collect all these results together, where the energy-dependent timescales are evaluated at a characteristic energy $E_p = 20$~TeV: 
\begin{eqnarray}\label{eq:timescales}
    &&t_\mathrm{acc}^{-1}\simeq  10^{-4}\;\frac{\sigma_{\rm tur}}{\eta}\frac{20 r_g}{M_7 R}\;\mathrm{s}^{-1} \, , \\  &&\nonumber t_\mathrm{esc}^{-1}\simeq 4.9\times 10^{-6}\;\eta^{2/3}\sigma^{-1/6}\;\left(\frac{20 r_g}{M_7 R}\right)^{-7/6}\;\mathrm{s}^{-1} \, , \\ 
    &&\nonumber t_{p\gamma}^{-1}\simeq 3.9\times 10^{-5}\;\left(\frac{20 r_g}{M_7 R}\right)^2\; \mathrm{s}^{-1} \, , \\ 
    &&\nonumber t_{pp}^{-1}\simeq 1.8\times 10^{-5}\;\frac{20 r_g}{M_7 R}\;\mathrm{s}^{-1} \, , \\ 
    &&\nonumber t_{\mathrm{BH}}^{-1}\simeq 1.2 \times 10^{-4}\; \left(\frac{20 r_g}{M_7 R}\right)^2\;\mathrm{s}^{-1} \, , \\
    &&\nonumber t_{\mathrm{synch}}^{-1}\simeq 7.1\times 10^{-7} \sigma  \left(\frac{20 r_g}{M_7 R}\right)^{-1}\;\mathrm{s}^{-1} \, .
\end{eqnarray} 
Note that the acceleration timescale depends on the magnetization associated with the turbulent field alone, $\sigma_{\rm tur} = \eta_B \sigma$, while the cooling and escape timescales either do not depend on the magnetization or depend on the total magnetization $\sigma$.

To estimate the relevant ranges for these parameters, we first notice that $t_{\rm acc}$ cannot be much larger than $t_{pp}$, otherwise $pp$ losses, which are mostly flat in energy above a few GeV, would lead to a very soft proton spectrum, impeding acceleration to the required energies. By comparing $t_{\rm acc}$ and $t_{pp}$ in Eq.~(\ref{eq:timescales}), we obtain the constraint
\begin{equation}\label{eq:constraint_1}
    \frac{\sigma_{\rm tur}}{\eta}\gtrsim 0.18 \, .
\end{equation}

It is reasonable to assume that the magnetic field is controlled by the dynamics of the accretion flow. Therefore, in the vicinity of the black hole, we assume that the coherence length can be as small as the gravitational radius, $\eta \gtrsim r_g/R$. For a corona with radius $R= 20-40\;r_g$, then we obtain $\eta\gtrsim 0.03 - 0.05$. 
This constraint alone suggests values of $\sigma_{\rm tur}\gtrsim 0.01$. We will later find that this range is suggested independently also by the IceCube signal, whose normalization requires comparable values of $\sigma_{\rm tur}$.

A further consideration is that BH losses should become competitive with particle acceleration at around $20$~TeV in order to explain the position of the peak of the IceCube spectrum. From this argument, by equating $t_{\rm acc}$ and $t_{\mathrm{BH}}$ in Eq.~(\ref{eq:timescales}), it follows that
\begin{equation}\label{eq:constraint_2}
    \frac{\sigma_{\rm tur}}{\eta}\simeq \frac{20 r_g}{M_7 R} \, .
\end{equation}
Using the constraint ~(\ref{eq:constraint_1}) in Eq. ~(\ref{eq:constraint_2}) results in the size of the corona being bounded by the condition 
\begin{equation}
    R \lesssim \frac{100 r_g}{M_7} \, .
\end{equation} 
For the remainder of this study, we adopt the benchmark values $M_7=1$ and $R=20 r_g$.

\begin{figure}
    \centering
    \includegraphics[width=0.5\textwidth]{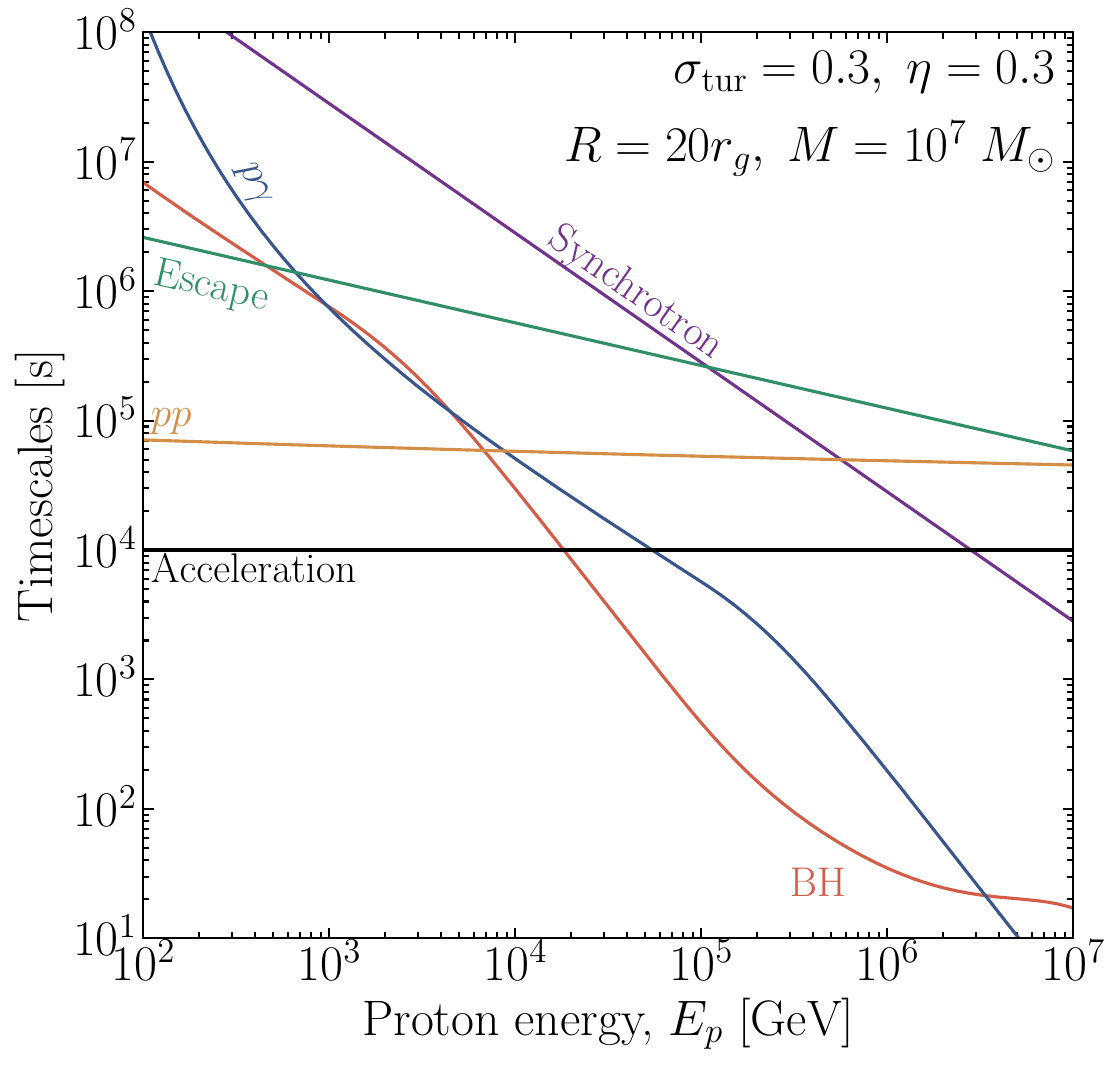}
    \caption{Energy-dependent timescales for the benchmark scenario considered in this work; numerical values of the parameters are shown in figure. We take $\eta_B=0.3$.}
    \label{fig:benchmark_timescales}
\end{figure}

Another essential constraint arises from the global energetics of the corona. Assuming the corona is magnetically powered,  the rate of magnetic energy dissipation will set the upper limit on its energetics.
This should be regarded as an upper limit on the energy available to energize the non-thermal protons producing the IceCube neutrino signal. We assume that the dissipation of the turbulent magnetic field $\delta B$ occurs on the magnetic reconnection timescale \citep{CS19}, which is given by 
\citep{Com24} \footnote{Here, we consider that the mean magnetic field acts as the non-reconnecting (guide) magnetic field component within magnetic reconnecting layers forming in the turbulent environment.}
\begin{eqnarray}
t_{\rm diss} \simeq \frac{1}{\eta_{\rm rec}} \sqrt{\frac{1+\sigma}{\sigma_{\rm tur}}} \frac{\ell}{c} \, ,
\end{eqnarray}
with $\eta_{\rm rec} \simeq 0.1$ \citep{ComissoJPP16,CassakJPP17}. Then, the rate of turbulent magnetic energy dissipation is 
\begin{eqnarray}
  &&L_B= \frac{1}{t_{\rm diss}} \frac{ \eta_B B^2}{8\pi} \frac{4\pi R^3}{3}\\ \nonumber &&\simeq  \frac{2\pi}{3} \frac{\eta_{\rm rec}}{\eta} \frac{c^3 \sigma_{\rm tur}^{3/2}}{(1+\sigma)^{1/2}}   {(n_p m_p+n_e m_e) R^2 } . 
\end{eqnarray}
Thus we find 
\begin{equation}
    L_B\simeq 2.1\times 10^{44}\;\frac{\sigma_{\rm tur}^{3/2}}{\eta\;\mathrm{max}(1,\sigma^{1/2})}\;\frac{R M_7}{20 r_g}\;\mathrm{erg/s}.
\end{equation}
Therefore, for $\sigma \leq 1$, the rate of dissipated magnetic energy depends on the magnetization associated with the turbulent field alone, $\sigma_{\rm tur} = \eta_B \sigma$, rather than on $\sigma$ and $\eta_B$ individually. Note that the rate of magnetic energy dissipation is comparable with the X-ray luminosity, which is consistent with the scenario of a magnetically-powered corona. However, this is not a required assumption of our subsequent analysis.

In Fig.~\ref{fig:benchmark_timescales} we show the energy-dependent timescales for the benchmark scenario we will consider throughout this work, with the parameters chosen so that the neutrino spectrum peaks in the correct energy range (see below). The main competition to the acceleration mechanism is BH energy losses, which dominate over acceleration above $20$~TeV. $p\gamma$ and synchrotron losses are subdominant in this energy range. $pp$ is about one order of magnitude slower than the acceleration timescale, and at $20$~TeV is comparable with the $p\gamma$ energy loss. The timescale for proton escape is much longer than the other timescales, and it does not strongly affect the proton spectrum around $20$~TeV. Nevertheless, as we will see, it plays a crucial role in shaping the proton spectrum at lower energies.

At this stage, we observe two crucial \emph{opposing tendencies}: the magnetization associated with the turbulent field of the corona cannot be too low, because otherwise the rate of dissipated magnetic energy cannot account for the observed neutrino luminosity, but neither it can be too large, otherwise BH losses are not efficient enough to counteract proton acceleration (see Eq.~(\ref{eq:timescales})) and fail to produce a peak in the neutrino spectrum at approximately $1$~TeV. These observations concur to point to the range 
\begin{equation}
0.01\lesssim \sigma_{\rm tur} \lesssim 1 
\end{equation}
as the relevant one for the corona of NGC~1068.

\section{Cosmic-ray transport and neutrino production}\label{sec:cosmic_ray_transport}

\begin{figure*}
    \centering
    \includegraphics[width=\textwidth]{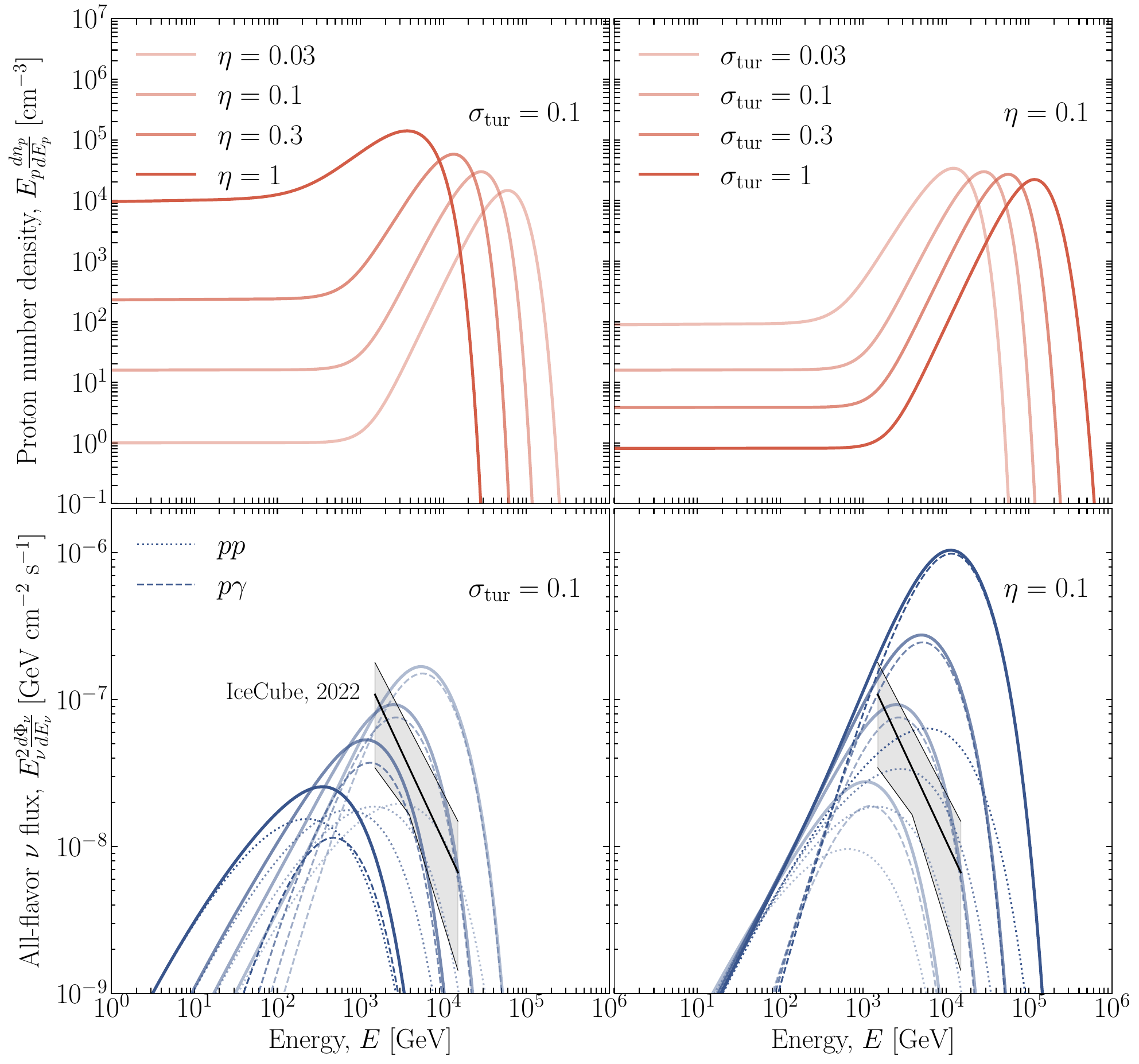}
    \caption{Proton number density (top) and all-flavor neutrino flux (bottom) for varying values of $\eta$ (left) and $\sigma_{\rm tur}$ (right) spaced from $0.03$ to $1$ (top and bottom). Results are obtained with $\eta_B=1$ and $\mathcal{F}=1$ (to highlight the maximum signal that can be achieved).}
    \label{fig:proton_density}
\end{figure*}

In this section, we discuss the dynamics of proton acceleration and cooling beyond the order-of-magnitude estimates presented in the previous section, and compute the neutrino spectrum produced in the corona. Our discussion is phrased in terms of a population of protons, which we describe by their phase-space distribution $f_p(p)$, where $p$ is their relativistic momentum, so that the number density of protons per unit energy is $f_p(p)4\pi p^2 dp$. Thus, the usual proton distribution per unit volume and energy is
\begin{equation}
    \frac{dn_p}{dE_p}=\frac{4\pi p^2}{c} f_p(p) |_{pc=E_p}
\end{equation}
for ultra-relativistic protons.

In our model, protons are accelerated over the timescale $t_\mathrm{acc}$ by the magnetized turbulence. This mechanism entails all non-thermal particles being accelerated at the same rate through scattering off large-scale turbulent fluctuations. A natural question is whether this process is indeed the most relevant in establishing a population of non-thermal particles. There are at least two aspects that are not captured by this model.

Firstly, there are rare scattering events which result in large energy gains, much larger than the mean squared momentum change $\Delta p^2 \simeq \sigma_{\rm tur} p^2$. These events lead to faster acceleration for a minority of particles, but their rate is determined by the amount of power in turbulent structures with very large velocity gradients. In Gaussian turbulence models, this power is exponentially suppressed. In realistic settings, the suppression is only of the power-law type, due to the intermittent formation of structures with large gradients. Therefore, low-energy particles are initially accelerated at a faster rate due to rare scattering events occurring on these structures with large gradients, with a typical spectrum $f_p(p)\propto p^{-4}$~\citep{Lemoine:2022rpj}.

Secondly, particles can also be accelerated in reconnecting current sheets~\citep{CS18,CS19,NB2021,Com2022}. If the guide field in the reconnection layers is not much larger than the reconnecting magnetic field, these particles undergo acceleration over a much shorter timescale, $t_{\rm rec} \simeq (\rho_g/c)/(\eta_{\rm B}^{1/2} \eta_{\rm rec})$ \citep{CS18,CS19}. However,  for the magnetization values relevant to this work ($\sigma_{\rm tur} \lesssim 1$), reconnection typically results in a soft spectrum $f_p(p)\propto p^{-s}$ with $s\geq 4-5$ (depending on the guide field). Once these particles escape the reconnection region, they enter the turbulent region; if their energy exceeds the threshold at which cooling dominates over turbulent acceleration, they would quickly be cooled down to lower energy; instead, if their energy is below this threshold, they enter with a spectrum steeper than the ones accelerated in regular turbulence, which as we will see typically predicts $f_p(p)\propto p^{-s}$ with $s \sim 3$. Hence, we will neglect the contribution of reconnection to the acceleration of high-energy non-thermal protons. We stress that this is true only in environments with relatively low $\sigma_p=B^2/4\pi n_p m_p c^2$, whereas in the scenario proposed in~\cite{Fiorillo:2023dts} we considered a very large value of $\sigma_p$, making reconnection the dominant acceleration mechanism.

Scattering on typical magnetic inhomogeneities, on the other hand, is a slower acceleration process, but it acts equally on all non-thermal particles. In a steady state, this mechanism becomes the most significant, as it tends to generate much harder spectra by energizing all non-thermal particles with the same efficiency. For this reason, here we focus on the simplest scenario in which only the mean energy gain resulting from turbulent stochastic acceleration is accounted for. The intermittency associated with the turbulent fluctuations, as well as the role of reconnection, is neglected. This assumption allows us to treat the turbulent acceleration using a Fokker-Planck approach, because rare events with large energization are not considered. Thus, the evolution of $f_p(p)$ can be written as
\begin{equation}\label{eq:Fokker_Planck}
   \frac{ \partial {f_p}}{{\partial t}}=\frac{1}{p^2}\frac{\partial}{\partial p}\left[\frac{p^4}{t_\mathrm{acc}}\frac{\partial f_p}{\partial p}\right]+\frac{1}{p^2}\frac{\partial}{\partial p}\left[\frac{p^3}{t_\mathrm{cool}(p)}f_p\right]-\frac{f_p}{t_\mathrm{esc}}+q_p(p).
\end{equation}
On the right-hand side, the first term accounts for the diffusive acceleration over the timescale $t_\mathrm{acc}$. This form of the acceleration term is analogous to the one conventionally used for gyroresonant mechanisms (see, e.g.,~\cite{Berezinsky:1990qxi} for a derivation), and was indeed already employed in~\cite{Murase:2019vdl} within this context. However, in the gyroresonant framework the acceleration timescale is energy-dependent, as particle acceleration depends on the turbulent power at a scale corresponding to the energy-dependent gyroradius. 
In contrast, first-principles PIC simulations have shown that the energization of particles in strong turbulence is dominated by nonresonant interactions with large-scale turbulent structures \citep{CS19,2020ApJ...893L...7W,2022PhRvD.106b3028B}, so the acceleration timescale is energy-independent and given by Eq.~(\ref{eq:acceleration_timescale}).

The second term on the right-hand side of Eq.~(\ref{eq:Fokker_Planck}) accounts for particle cooling, characterized by the timescale 
\begin{equation} 
t_\mathrm{cool}^{-1}=t_{p\gamma}^{-1}+t_\mathrm{BH}^{-1}+t_{pp}^{-1}+t_\mathrm{synch}^{-1}.
\end{equation}
The third term describes the escape of particles, and the last term, $q_p(p)$, describes the low-energy injection of protons in the regions of turbulent acceleration, which comes, e.g., from the pre-acceleration by reconnection layers within the turbulent cascade. The specific form of $q_p(p)$ is inessential, since it is only relevant at very low energies, much smaller than the ones to which protons are accelerated.

We are interested in the stationary solution of Eq.~(\ref{eq:Fokker_Planck}). Thus, we solve it without the time derivative. The normalization of the solution would be determined by the low-energy source term $q_p(p)$, which however is not directly accessible by observations. However, we can relate the normalization to the rate with which the magnetic turbulence energizes the non-thermal particle spectrum. From Eq.~(\ref{eq:Fokker_Planck}), we have that the amount of energy transferred to the non-thermal protons per unit time and volume is
\begin{equation}
    Q_p=\frac{dU_p}{dtdV}=-4\pi \int \frac{p^4}{t_{\rm acc}}\frac{\partial f_p}{\partial p}dp,
\end{equation}
obtained by integrating the acceleration term in Eq.~(\ref{eq:Fokker_Planck}), multiplied by $p^3 dp$, over all energies.
Thus, we can normalize the proton spectrum such that the energy flowing to the protons is a fixed fraction of the rate of magnetic energy dissipation, 
\begin{equation}
    Q_p=\mathcal{F}\frac{L_B}{\frac{4}{3}\pi R^3}.
\end{equation}
Below, we will show results for the extreme case $\mathcal{F}=1$, to show the maximum proton and neutrino production rates that could be achieved by turbulence.

Fig.~\ref{fig:proton_density} shows the stationary proton spectrum obtained in this scenario, for varying values of $\sigma_{\rm tur}$ and $\eta$. At high energies, as expected, the spectrum is rapidly suppressed by the fast cooling. Lower values of $\sigma_{\rm tur}$ and higher values of $\eta$ move this transition at lower energies, since they lead to slower acceleration. At low energies, the dominant process shaping the spectrum is the acceleration. 
One easily verifies that, in this regime, Eq.~(\ref{eq:Fokker_Planck}) in steady-state admits two independent solutions, $f_p\propto p^{-3}$ and $f_p\propto p^0$. 
Below the peak, the solution transitions from the former (at low energies) to the latter (close to the peak), which shows as a ``pile-up'' region. The position of the transition is determined by the rate of escape, despite the latter being slower than any other process. Indeed, in the absence of escape, Eq.~(\ref{eq:Fokker_Planck}) does not admit any steady solution, since the number of particles would be perpetually increased by the injection term with no process countering it. The steady solution we find arises from the balance between injection and the slow escape.
Note that in the gyroresonant framework, the spectral index in the low-energy region would be highly sensitive to the turbulence spectrum. Instead, since the interaction between protons and turbulence is non-resonant, the behavior of the distribution in this region turns out to be universal. For neutrinos, the production is dominated by the peak region, and therefore the timescale for escape (which determines the break in the proton spectrum) does not play a substantial role in shaping the neutrino spectrum.

With the proton distribution in hand, we can now determine the neutrino spectrum steadily produced in the corona. To achieve this, we compute the neutrinos produced from the decay of pions originating in $pp$ and $p\gamma$ collisions. To calculate the spectra of the produced neutrinos, we adopt the fit functions provided in~\cite{Kelner:2006tc} and~\cite{Kelner:2008ke}. For $p\gamma$ interactions, we consider as a target the photon spectrum from Fig.~\ref{fig:photon_spectrum}. We assume a distance between the source and the Earth of $d=10.1$~Mpc~\citep{Tully_2008,Padovani2024}.

The bottom panels of Fig.~\ref{fig:proton_density} show the all-flavor neutrino flux at Earth, compared with the neutrino signal observed by IceCube assuming flavor equipartition at Earth (thus we multiply by $3$ the IceCube muon neutrino flux). The neutrino signal is dominated by a competition of $pp$ and $p\gamma$ interactions. $p\gamma$ interactions produce a harder neutrino spectrum, due to the photohadronic efficiency increasing with energy, and dominate close to the peak of the spectrum, while $pp$ production dominates at lower energies. 
$pp$ interactions lead to a neutrino flux $d\Phi_\nu/dE_\nu\propto E_\nu^{-1}$ -- the pile-up features in the proton spectrum are mostly washed out in the neutrino production and the neutrino spectrum tracks the proton spectrum $f_p\propto p^{-3}$ below the break. The $p\gamma$ neutrino spectrum is hardened as $d\Phi_\nu/dE_\nu\propto E_\nu^{-2+s_\gamma}$ (see, e.g., \cite{Winter:2012xq,Fiorillo:2021hty}), where $s_\gamma$ is the local spectral index of the photons $n_\gamma(E_\gamma)\propto E_\gamma^{-s_\gamma}$, which in the X-ray coronal spectrum is close to $s_\gamma\simeq 2$. As $\eta$ increases, the acceleration rate decreases, causing the neutrino spectrum to peak at lower energies. On the contrary, increasing $\sigma_{\rm tur}$ tends to increase the acceleration rate. Notice that even with $\eta=0.1$, $\sigma_{\rm tur}=0.03$ is unable to match the IceCube signal, so the range $\sigma_{\rm tur}\gtrsim 0.01$ again emerges as a natural requisite to fit the normalization of the IceCube spectrum. In turn, for $\sigma_{\rm tur}=0.01$, the coherence length, and therefore $\eta$, cannot be too large, otherwise acceleration becomes too slow and the neutrino signal is underpredicted compared to the IceCube measurement.

\begin{figure}
    \centering
    \includegraphics[width=0.5\textwidth]{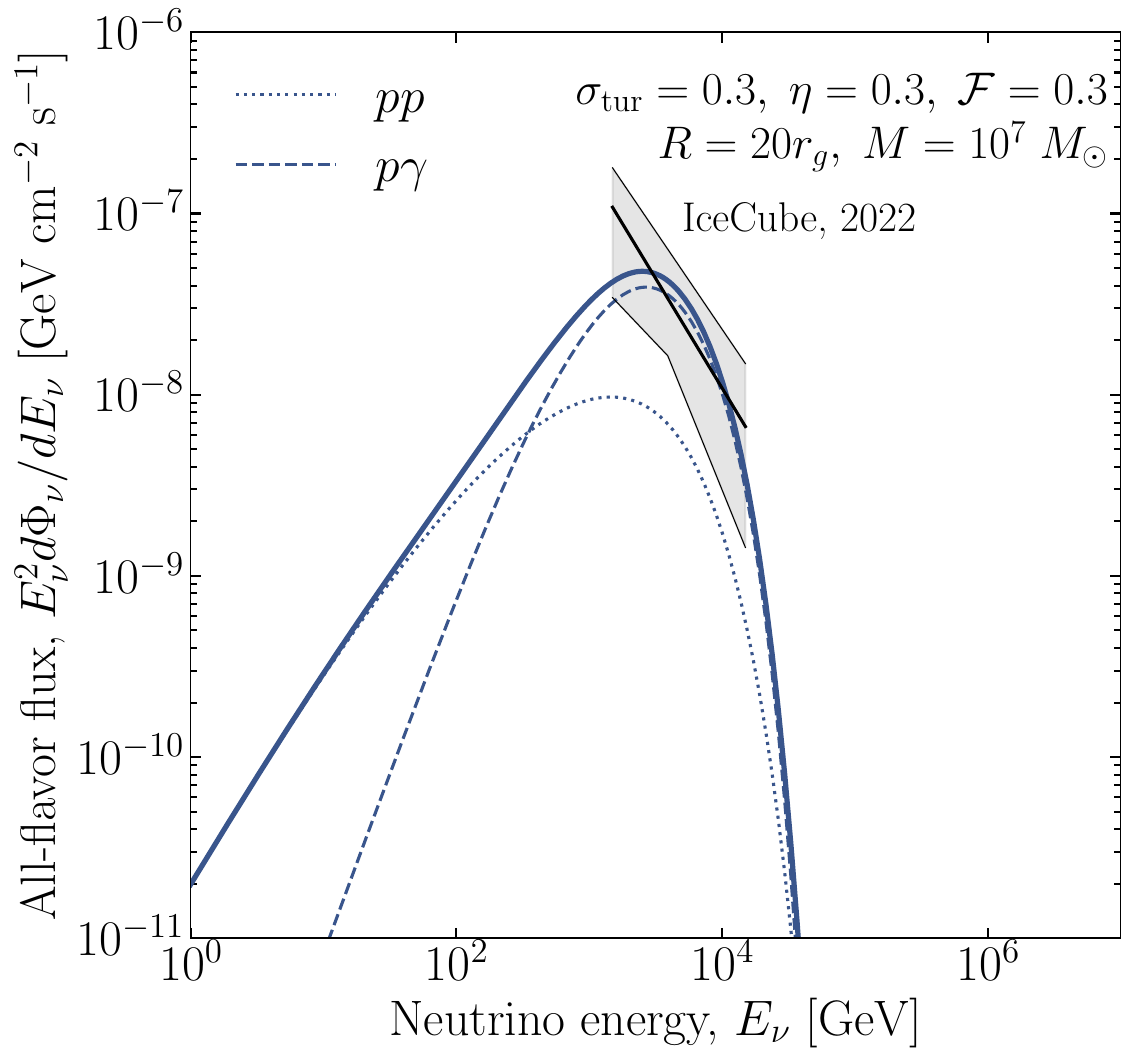}
    \caption{Neutrino spectrum, separated in its $pp$ and $p\gamma$ components, for the benchmark case $\sigma_{\rm tur}=0.3$ and $\eta=0.3$; we choose the normalization with a factor $\mathcal{F}=0.3$ to fit the IceCube observation.}
    \label{fig:neutrino_benchmark}
\end{figure}

As a benchmark case, we now focus on the choice $\sigma_{\rm tur}=\eta=0.3$ (we assume again $\eta_B=0.3$, therefore $\sigma=1$). For the adopted benchmark coronal radius, the proton plasma beta is $\beta\simeq 0.03$.
For this case, it is now helpful to quantify the main global properties of the proton population. We consider now a fraction $\mathcal{F}=0.3$, which agrees with the IceCube observations, as shown in Fig.~\ref{fig:neutrino_benchmark}. 
The total number density of non-thermal protons, after integrating over energy, is $n_p^\mathrm{nt}\simeq 1.9\times 10^4$~cm$^{-3}$, and as we can see from Fig.~\ref{fig:proton_density}, is mostly dominated by protons at the peak energy, close to the pile-up region. This should be compared with the density of thermal protons in the corona $n_p\simeq 2.5\times 10^{10}$~cm$^{-3}$, which is about six orders of magnitude higher.

While a comprehensive assessment of the non-thermal to thermal proton ratio is beyond the scope of the present work, a few observations can be made. In the large-amplitude turbulence scenario under consideration, one can estimate the fraction of injected protons by observing that within one $\ell/c$, the volume of plasma going through the reconnecting current sheets is given by \citep{CS19} $V_{\rm rec} \sim \eta_{\rm rec} 4 \pi R^3/3$. Furthermore, for current sheets with a guide magnetic field comparable to the reconnecting component, as expected when $\delta B/B \sim 1/2$, the typical power-law slope for protons with $p \geq p_0$ is $dN_p/dpdV \propto p^{-[3,4]}$ \citep{Com24}, where $p_0$ is the characteristic momentum of the protons processed by reconnection. If protons start to be accelerated by scattering off large-scale turbulent fluctuations at an injection momentum of order $p_{\rm inj} \sim 10^2 p_0$, then current sheets inject a fraction $\sim 10^{-4} - 10^{-6}$ of the processed protons. Combining this with the fraction of protons processed by reconnection, the overall fraction of thermal protons injected into turbulent acceleration is $\sim 10^{-4}\eta_{\rm rec} - 10^{-6}\eta_{\rm rec} \sim 10^{-5} - 10^{-7}$, in accordance with the normalization adopted for the IceCube neutrino signal.

The energy density of the non-thermal proton population is $u_p\simeq 8\times 10^5$~erg/cm$^3$. For comparison, the magnetic field energy density of the turbulent field for our benchmark scenario is $u_B=\sigma_{\rm tur} n_p m_p c^2/2\simeq 1.9\times 10^7$~erg/cm$^3$. 
Hence, protons have an energy density more than an order of magnitude lower than the magnetic field energy density, suggesting a negligible impact of proton acceleration on the turbulent cascade. On the other hand, if the non-thermal protons were to draw an energy comparable to that of the turbulence fluctuations, damping effects on the turbulent cascade might become significant \citep{Lemoine24},  potentially resulting in a steepening of the proton energy spectrum.

\begin{figure}
    \centering
    \includegraphics[width=0.5\textwidth]{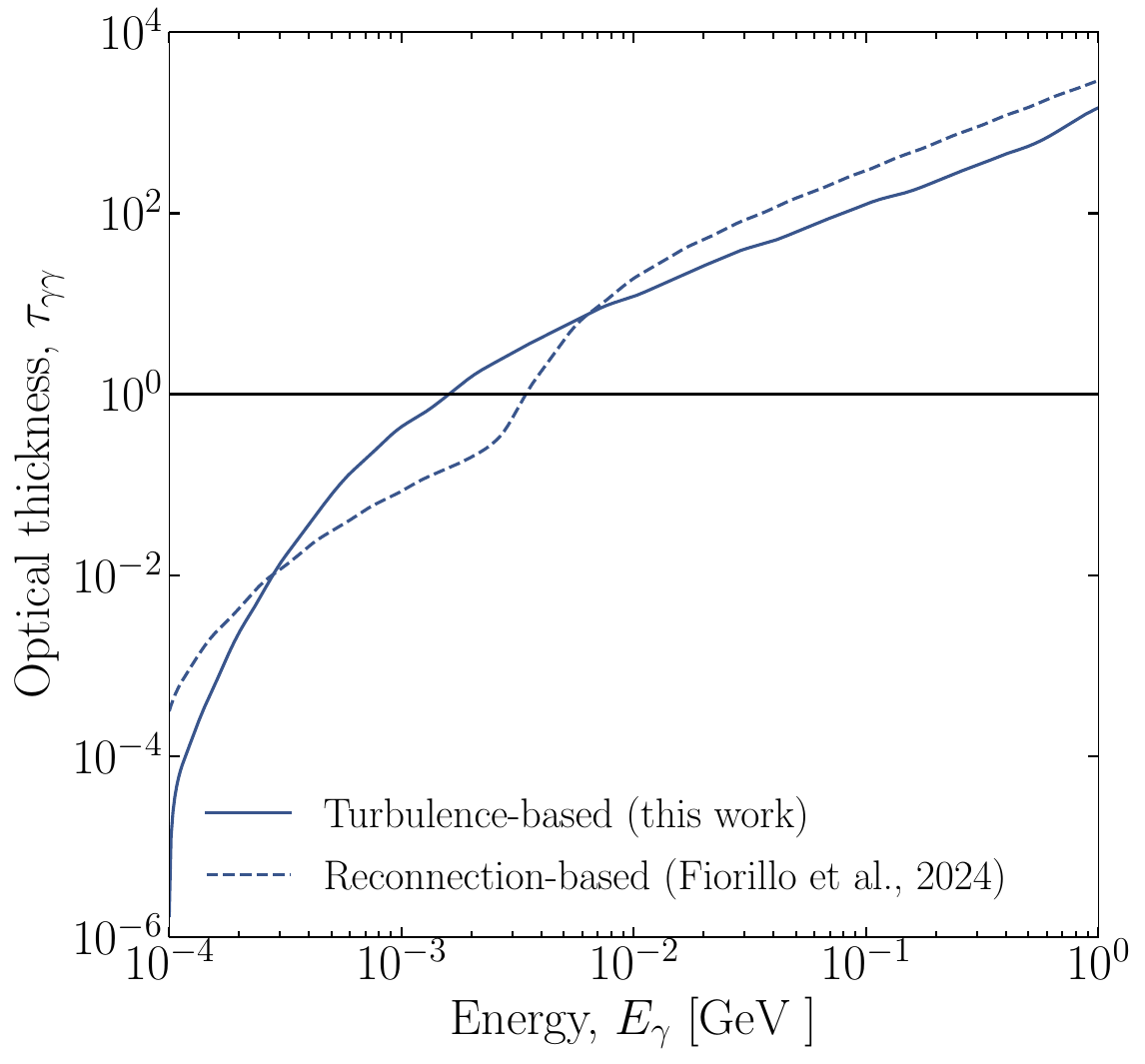}
    \caption{Optical thickness for photon-photon interactions in the corona from the electromagnetic spectrum in the turbulence-based model (from Fig.~\ref{fig:photon_spectrum}) and in the reconnection-based model (from Fig.~3 of~\cite{Fiorillo:2023dts}).}
    \label{fig:optical_thickness}
\end{figure}

Finally, let us comment on the possible electromagnetic signatures. $pp$ and $p\gamma$ interactions produce $\pi^0$, which decay to photons subsequently interacting via $\gamma\gamma$ scattering with the photon field and producing an electromagnetic cascade. In addition, BH interactions produce $e^+e^-$ which emit photons via synchrotron and Compton radiation, contributing to the cascade. For this work, we do not delve into a full computation of the output of this cascade as we did in~\cite{Fiorillo:2023dts}, since the neutrino signal by itself is already informative on the coronal properties. However, we verify explicitly that the corona is sufficiently optically thick to $\gamma\gamma$ interactions to absorb photons in the GeV range, impeding them from exceeding the upper limits from MAGIC~\citep{MAGIC-UL-NGC1068}.

To show this, we determine the optical thickness $\tau_{\gamma\gamma}$ for a photon with energy $E_\gamma$ traveling through the corona of radius $R$. For the interaction cross section of $\gamma\gamma$ pair production, we use the approximation presented in Eq.~(4.7) of~\cite{1990MNRAS.245..453C}. For the reconnection-based model, we use the benchmark size of the corona in the spherical model of~\cite{Fiorillo:2023dts}, namely $R=1.4\times 10^{12}$~cm. Fig.~\ref{fig:optical_thickness} shows the optical thickness $\tau_{\gamma\gamma}$ in both scenarios. In the reconnection-based model, the corona is much more compact, leading to a much denser target of photons. However, in the turbulence-based model the assumed photon luminosity is higher by about an order of magnitude, which seems to be somewhat required for BH losses to compete with acceleration at the relevant energies. The two factors cancel out, so that overall the optical thickness for the two models is pretty similar. In both cases, photons above about $10$~MeV are strongly attenuated.

\section{Discussion}

Magnetized turbulence is one of the most compelling candidates as a mechanism for accelerating the protons responsible for the neutrino signal originating from NGC~1068. We have shown that this scenario is feasible if strong turbulence ($\delta B \sim B$) with magnetization in the range of $0.01 \lesssim \sigma_{\rm tur} \lesssim 1$ characterize the corona, since otherwise the rate of dissipated magnetic energy is insufficient to explain the normalization of the IceCube observed flux. In this regime, the often-used assumption that the interaction between cosmic rays and magnetized turbulence is gyroresonant breaks down, since gyroresonances are strongly broadened, and non-resonant interactions dominate. One can still define a diffusion coefficient for the proton energy, but the corresponding acceleration timescale $t_{\rm acc}$, namely the typical time for the acceleration of non-thermal protons, is essentially energy-independent, as inferred from PIC simulations~\citep{CS19,2020ApJ...893L...7W,Zhdankin20,2022PhRvD.106b3028B}. This contrasts sharply with the gyroresonant model, where $t_{\rm acc}$ depends on the amount of power in the turbulent cascade at scales resonant with the gyroradius of the particle. On the other hand, in the more appropriate non-resonant model, the largest scales of the turbulent cascade control the acceleration rate of all particles.

From these two main changes -- relatively large $\sigma_{\rm tur}$ and acceleration timescale independent of energy -- follow significant consequences. The rate of acceleration is much faster than in the scenarios with gyroresonant stochastic acceleration considered previously~\citep{Murase:2019vdl}. Therefore, in order to explain the soft IceCube neutrino spectrum, BH cooling must effectively compete with the stochastic acceleration of non-thermal protons, which in turn requires more compact coronae compared to prior studies. The typical timescale at which the proton spectrum evolves is of the order of $10^4$~s. Note that the turbulent scenario envisioned here is truly stationary, with turbulence sustained steadily by energy injection at large scales. In this sense, our model differs from the turbulent acceleration considered in~\cite{Mbarek:2023yeq}, which concluded that turbulence would produce a spectrum $f_p\propto p^{-5}$. This conclusion is appropriate only for the case of transient turbulence, in which particles are rapidly accelerated by intermittent fluctuations. In contrast, if turbulence is maintained in a steady state, as in our case, protons are accelerated into a much harder spectrum with $f_p\propto p^{-3}$.

Due to the compactness of the corona, our scenario of turbulence-driven acceleration does not lead to a sizable gamma-ray flux at GeV energies. Thus, neutrinos are the only messenger that at present can pinpoint some of the properties of the non-thermal protons. We leave a more detailed analysis of the gamma-ray emission that would arise from the cascade of gamma-rays injected by photohadronic interactions for future work. Our findings have shown that the neutrino signal by itself is already sufficient to constrain the allowed range for $\sigma_{\rm tur}$, $\eta$, and the corona size.

Our deduction of a more compact corona seems required to ensure that the timescale of BH losses is comparable to the short particle acceleration timescale. This is found in agreement with model-independent multi-messenger constraints discussed in \cite{Murase_2022ApJL}.
One might consider reducing the acceleration rate by decreasing $\sigma_{\rm tur}$, which could potentially lead to scenarios with a larger corona (e.g.~\cite{Murase:2019vdl}). However, crucially, lowering $\sigma_{\rm tur}$ slows down proton acceleration, making it easier to accommodate the peak energy of the neutrino spectrum, but also lowers the rate of dissipated magnetic energy, reducing the energy budget available for neutrino production. For example, for a magnetic field $B\simeq 1$~kG and a turbulence strength $\eta_B\simeq 0.01$, the Alfvèn velocity associated with the turbulent field is $v_A^2/c^2=\sigma_{\rm tur}=2\eta_B u_B/n_p m_p c^2\simeq 5\times 10^{-5}$ with $n_p=10^{10}$~cm$^{-3}$. Hence, for a coronal volume $V\simeq 10^{43}$~cm$^3$, similarly to what was determined in some prior studies, the rate of dissipated magnetic energy, assuming a turbulence scale of $\ell=10^{14}$~cm, is $L_B=\eta_B u_B V (v_A/\ell)\simeq 9\times 10^{39}$~erg/s, two orders of magnitude lower than the observed neutrino luminosity, and therefore not sufficient to power the required non-thermal proton energization.

The scenario we consider here is also entirely distinct from our previous proposal of reconnection-driven proton acceleration outlined in~\cite{Fiorillo:2023dts}. In that study, we considered a much more compact corona in the black hole magnetosphere with very high $\sigma_p=B^2/4\pi n_p m_p c^2$ (low-density region). The acceleration mechanism in that case is entirely due to magnetic reconnection. Additionally, the reconnection-based scenario is likely to result in multiple flaring events, each associated with the formation of a transient reconnection layer, whose typical lifetime is $100$~s. Instead, in the present scenario, turbulence is for all practical purposes stationary. It is an intriguing question whether future measurements can truly disentangle the two scenarios of particle acceleration driven by reconnection or turbulence. The transient nature of the reconnection-based picture would certainly be a useful observational test; while it is unlikely that neutrino measurements could exhibit these transient features, future MeV telescopes would be sensitive to the photons from the electromagnetic cascade which would exhibit a similar temporal variability. Another element of difference, which is perhaps the most interesting for future measurements of the neutrino signal at IceCube-Gen2, is the neutrino spectrum; in the reconnection-based scenario, the signal has a power-law spectrum with $f_p\propto p^{-5}$ (for strong guide fields, as we hypothesized in~\cite{Fiorillo:2023dts}), whereas in the turbulence-based scenario considered here cooling eventually inhibits particle acceleration and therefore the spectrum cuts off exponentially. 
Overall, despite operating on fundamentally different physical principles compared to the reconnection-based model, the turbulence-based scenario convincingly accounts for the main features of the neutrino signal observed at IceCube.

\section*{Acknowledgements} 
We are grateful to Martin Lemoine, Kohta Murase, and Philipp Kempski for useful conversations. 
D.F.G.F. is supported by the Villum Fonden under Project No.\ 29388 and the European Union's Horizon 2020 Research and Innovation Program under the Marie Sk{\l}odowska-Curie Grant Agreement No.\ 847523 ``INTERACTIONS.'' 
L.C is supported by the NSF grant PHY-2308944 and the NASA ATP grant 80NSSC22K0667.
E.P. acknowledges support from the Agence Nationale de la Recherche (grant ANR-21-CE31-0028).
M.P. acknowledges support from the Hellenic Foundation for Research and Innovation (H.F.R.I.) under the ``2nd call for H.F.R.I. Research Projects to support Faculty members and Researchers'' through the project UNTRAPHOB (ID 3013). 
L.S. acknowledges support from the Simons Foundation grant 00001470, from the DoE Early Career Award DE-SC0023015, and from the NSF grant PHY-2206609.

\bibliographystyle{aasjournal}
\bibliography{References}

\end{document}